\documentclass[aps, prl, twocolumn, superscriptaddress]{revtex4-1}
\usepackage{amsmath, graphicx, float, upgreek, hyperref, mathpazo, bm}
\usepackage[T1]{fontenc}
\usepackage{soul, color,upgreek}
\hypersetup{colorlinks=true, citecolor=blue, urlcolor=blue}

\begin{document}

\title{\huge Deep Learning with Coherent Nanophotonic Circuits}
\author{\textsf{Yichen Shen$^{1*}$, Nicholas C. Harris$^{1*}$, Scott Skirlo$^{1}$, Mihika Prabhu$^{1}$, Tom Baehr-Jones$^{2}$, Michael Hochberg$^{2}$, Xin Sun$^{3}$, Shijie Zhao$^{4}$, Hugo Larochelle$^{5}$, Dirk Englund$^{1}$, and Marin Solja\v{c}i\'{c}}}
\affiliation{
\small{Research Laboratory of Electronics, Massachusetts Institute of Technology, Cambridge, MA 02139, USA}\\
\small{$^{2}$Coriant Advanced Technology, 171 Madison Avenue, Suite 1100, New York, NY 10016, USA}\\
\small{$^{3}$Department of Mathematics, Massachusetts Institute of Technology, Cambridge, MA 02139, USA}\\
\small{$^{4}$Department of Biology, Massachusetts Institute of Technology, Cambridge, MA 02139, USA}\\
\small{$^{5}$Twitter Inc., 141 Portland St, Cambridge, MA 02139, USA}\\
\small{$^\star$\textbf{These authors contributed equally to this work.}} \\
}

\begin{abstract}
Artificial Neural Networks are computational network models inspired by signal processing in the brain. These models have dramatically improved the performance of many learning tasks, including speech and object recognition. However, today's computing hardware is inefficient at implementing neural networks, in large part because much of it was designed for von Neumann computing schemes. Significant effort has been made to develop electronic architectures tuned to implement artificial neural networks that improve upon both computational speed and energy efficiency. Here, we propose a new architecture for a fully-optical neural network that, using unique advantages of optics, promises a computational speed enhancement of at least two orders of magnitude over the state-of-the-art and three orders of magnitude in power efficiency for conventional learning tasks. We experimentally demonstrate essential parts of our architecture using a programmable nanophotonic processor.
\end{abstract}

\maketitle

\begin{figure*}[ht]
\begin{center}
\includegraphics[width=4.2in]{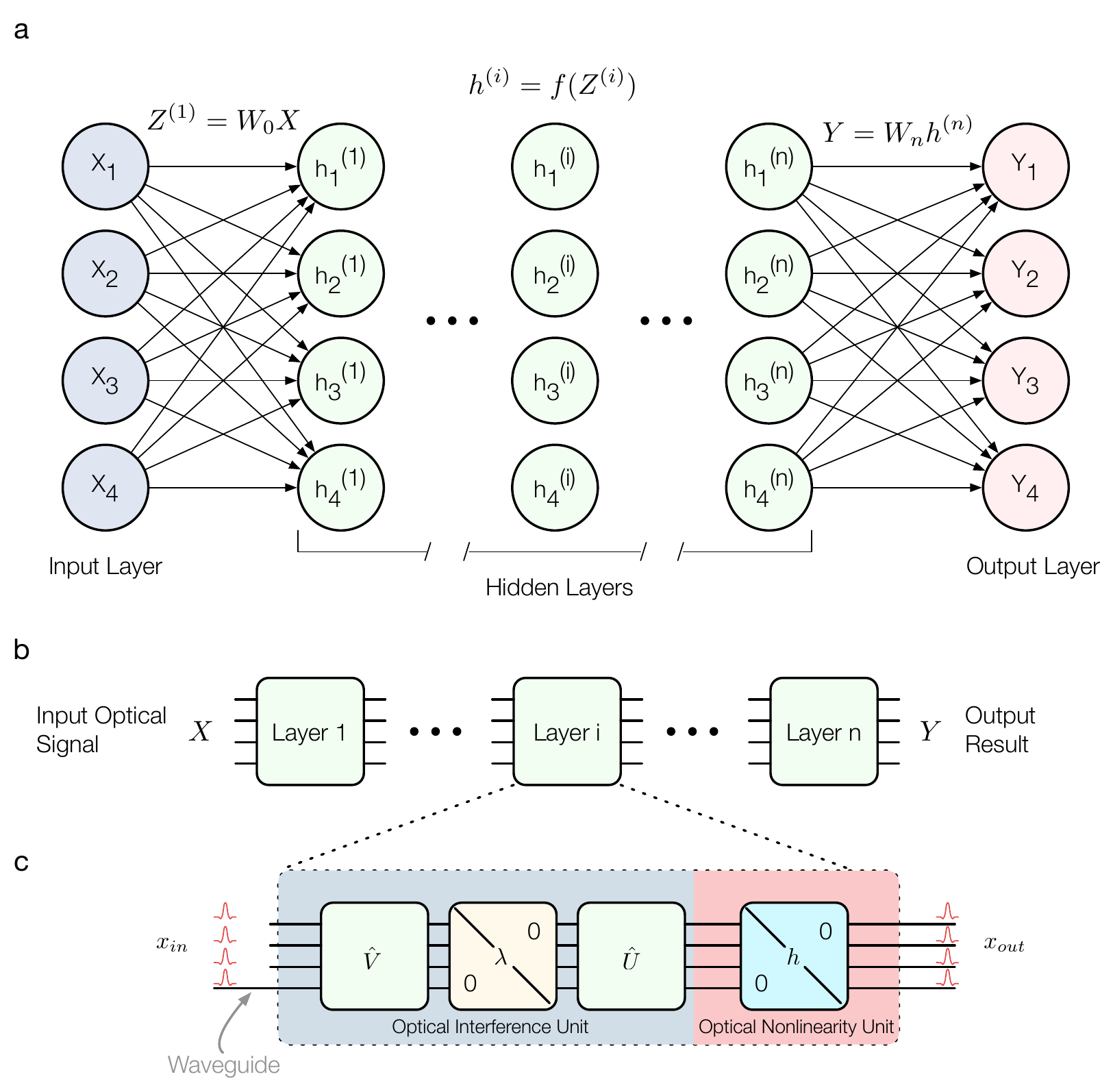}
\caption{\textbf{General Architecture of Optical Neural Network}
a. General artificial neural network architecture composed of an input layer, a number of hidden layers, and an output layer.
b. Decomposition of the general neural network into individual layers.
c. Optical interference and nonlinearity units that compose each layer of the artificial neural network.
}
\label{fig:general_architecture}
\end{center}
\end{figure*}

Modern computers based on the von Neumann architecture are far more power-hungry and less effective than their biological counterparts -- central nervous systems -- for a wide range of tasks including perception, communication, learning, and decision making. With the increasing data volume associated with processing \textit{big data}, developing computers that learn, combine, and analyze vast amounts of information quickly and efficiently is becoming increasingly important. For example, speech recognition software (e.g., Apple's Siri) is typically executed in the cloud since these computations are too taxing for mobile hardware; real-time image processing is an even more demanding task \cite{hinton2006reducing}. To address the shortcomings of von Neumann computing architectures for neural networks, much recent work has focused on increasing artificial neural network computing speed and power efficiency by developing electronic architectures (such as ASIC and FPGA chips) specifically tailored to a task~\cite{mead1990neuromorphic,10.3389/fnins.2011.00108,shafiee2016isaac,misra2010artificial}. Recent demonstrations of electronic neuromorphic hardware architectures have reported improved computational performance~\cite{silver2016mastering}. Hybrid optical-electronic systems that implement spike processing \cite{tait2014photonic,tait2014broadcast,prucnal2016recent} and reservoir computing \cite{vandoorne2014experimental,appeltant2011information} have also been investigated recently. However, the computational speed and power efficiency achieved with these hardware architectures are still limited by electronic clock rates and ohmic losses.

Fully-optical neural networks offer a promising alternative approach to microelectronic and hybrid optical-electronic implementations. Linear transformations (and certain non-linear transformations) can be performed at the speed of light and detected at rates exceeding 100~GHz~\cite{Vivien:12} in photonic networks, and in some cases, with minimal power consumption \cite{Cardenas2009}. For example, it is well known that a common lens performs Fourier transform without any power consumption, and that certain matrix operations can also be performed optically without consuming power. However, implementing such transformations with bulk optical components (such as fibers and lenses) has been a major barrier because of the need for phase stability and large neuron counts. Integrated photonics solves this problem by providing a scalable solution to large, phase-stable optical transformations~\cite{Harris:2015ux}.

Here, we experimentally demonstrate on-chip, coherent, optical neuromorphic computing on a vowel recognition dataset. We achieve a level of accuracy comparable to a conventional digital computer using a fully connected neural network algorithm. We show that, under certain conditions, the optical neural network architecture can be at least two orders of magnitude faster for forward propagation while providing linear scaling of neuron number versus power consumption. This feature is enabled largely by the fact that photonics can perform matrix multiplications, a major part of nerual network algorithms, with extreme energy efficiency. While implementing scalable von Neumann optical computers has proven challenging, artificial neural networks implemented in optics can leverage inherent properties, such their weak requirements on nonlinearities, to enable a practical, all-optical computing application. An optical neural network architecture can be substantially more energy efficient than conventional artificial neural networks implemented on current electronic computers.

\section{Optical Neural Network Device Architecture}

\begin{figure*}[ht]
\begin{center}
\includegraphics[width=\textwidth]{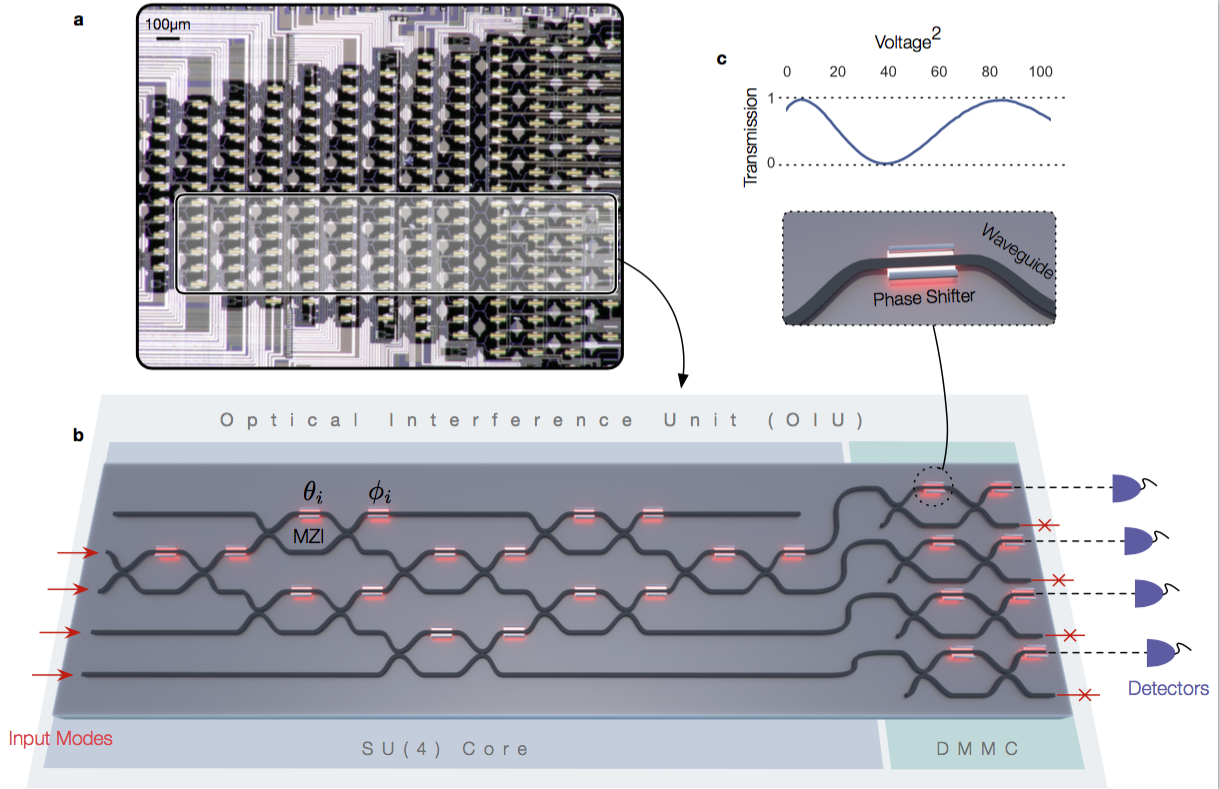}
\caption{\textbf{Illustration of Optical Interference Unit}
a. Optical micrograph of an experimentally fabricated 22-mode on-chip optical interference unit; the physical region where the optical neural network program exists is highlighted in grey. The system acts as an optical field-programmable gate array--a test bed for optical experiments.
b. Schematic illustration of the optical neural network program demonstrated here which realizes both matrix multiplication and amplification fully optically.
c. Schematic illustration of a single phase shifter in the Mach-Zehnder Interferometer (MZI) and the transmission curve for tuning the internal phase shifter of the MZI.
}
\label{fig:Optical_interference}
\end{center}
\end{figure*}

An artificial neural network (ANN)~\cite{lecun2015} consists of a set of input artificial neurons (represented as circles in Fig.~\ref{fig:general_architecture}(a)) connected to at least one hidden layer and an output layer. In each layer (depicted in Fig.~\ref{fig:general_architecture}(b)), information propagates by linear combination (e.g. matrix multiplication) followed by the application of a nonlinear activation function. ANNs can be trained by feeding training data into the input layer and then computing the output by forward propagation; weighting parameters in each matrix are subsequently optimized using back propagation~\cite{schmidhuber2015deep}.

The Optical Neural Network (ONN) architecture is depicted in Fig.~\ref{fig:general_architecture} (b,c). As shown in Fig.~\ref{fig:general_architecture}(c), signals are encoded in the amplitude of optical pulses propagating in integrated photonic waveguides where they pass through an optical interference unit (OIU) and finally an optical nonlinearity unit (ONU). Optical matrix multiplication is implemented with an OIU and nonlinear activation is realized with an ONU.

To realize an OIU that can implement any real-valued matrix, we use the singular value decomposition (SVD)~\cite{lawson1995solving} since a general, real-valued matrix ($\mathbf{M}$) may be decomposed as
$\mathbf{M}=\mathbf{U} \boldsymbol{\Sigma} \mathbf{V}^{*} $, where $U$ is an $m \times m$ unitary matrix, $\Sigma$ is a $m \times n$ diagonal matrix with non-negative real numbers on the diagonal, and $V^*$ is the complex conjugate of the $n \times n$ unitary matrix $V$. It was theoretically shown that any unitary transformations $\mathbf{U}, \mathbf{V^{*}}$ can be implemented with optical beamsplitters and phase shifters~\cite{PhysRevLett.73.58, Miller:15}. Matrix multiplication implemented in this manner consumes, in principle, no power. The fact that a major part of ANN calculations involves matrix products enables the extreme energy efficiency of the ONN architecture presented here. Finally, $\boldsymbol{\Sigma}$ can be implemented using optical attenuators; optical amplification materials such as semiconductors or dyes could also be used~\cite{connelly2007semiconductor}.

The ONU can be implemented using optical nonlinearities such as saturable absorption~\cite{selden1967pulse,Bao2010,schirmer1997nonlinear} and bistability~\cite{soljavcic2002optimal,PhysRevLett.71.3959,PhysRevB.62.R7683,Nozaki2010,Ros2015} that have been demonstrated seperately in photonic circuits. For an input intensity $I_{in}$, the optical output intensity is thus given by a nonlinear function $I_{out}=f(I_{in})$ ~\cite{NIPS2012_4824}.

\section{Experiment}
For an experimental demonstration of our ONN architecture, we implement a two layer, fully connected neural network with the OIU shown in Fig.~\ref{fig:Optical_interference} and use it to perform vowel recognition. To prepare the training and testing dataset, we use 360 datapoints that each consist of four log area ratio coefficients~\cite{chow2004speaker} of one phoneme. The log area ratio coefficients, or feature vectors, represent the power contained in different logarithmically-spaced frequency bands and are derived by computing the Fourier transform of the voice signal multiplied by a Hamming window function. The 360 datapoints were generated by 90 different people speaking 4 different vowel phonemes~\cite{deterding1990speaker}. We use half of these datapoints for training and the remaining half to test the performance of the trained ONN. We train the matrix parameters used in the ONN with the standard back propagation algorithm using stochastic gradient descent method \cite{hinton2006reducing}, on a conventional computer. Further details on the dataset and backpropagation procedure are included in Supplemental Information Section 3.

The coherent ONN is realized with a programmable nanophotonic processor~\cite{Harris:2015ux} composed of an array of 56 Mach-Zehnder interferometers (MZIs) and 213 phase shifting elements, as shown in Fig.~\ref{fig:Optical_interference}. Each interferometer is composed of two evanescent-mode waveguide couplers sandwiching an internal thermo-optic phase shifter~\cite{harris2014efficient} to control the splitting ratio of the output modes, followed by a second modulator to control the relative phase of the output modes. By controlling the phase imparted by these two phase shifters, these MZIs perform all rotations in the $SU(2)$ Lie group given a controlled incident phase on the two electromagnetic input modes of the MZI. The nanophotonic processor was fabricated in a silicon-on-insulator photonics platform with the OPSIS Foundry ~\cite{baehr201225}.

To experimentally realize arbitrary matrices by SVD, we programmed an $SU(4)$ core~\cite{PhysRevLett.73.58, miller2013selfconfiguring} and a non-unitary diagonal matrix multiplication core (DMMC) into the nanophotonic processor~\cite{Harris:2015ux,harris2014efficient}, as shown in Fig.~\ref{fig:Optical_interference} (b). The $SU(4)$ core implements operators $U$ and $V$ by a Givens rotations algorithm~\cite{PhysRevLett.73.58, miller2013selfconfiguring} that decomposes unitary matrices into sets of phase shifters and beam splitters, while the DMMC implements $\Sigma$ by controlling the splitting ratios of the DMMC interferometers to add or remove light from the optical mode relative to a baseline amplitude. The measured fidelity for the 720 OIU and DMMC cores used in the experiment was 99.8 $\pm$ 0.003 \%; see methods for further detail.

In this analog computer, fidelity is limited by practical non-idealities such as (1) finite precision with which an optical phase can be set using our custom 240-channel voltage supply with 16-bit voltage resolution per channel (2) photodetection noise, and (3) thermal cross-talk between phase shifters which effectively reduces the number of bits of resolution for setting phases. As with digital floating-point computations, values are represented to some number of bits of precision, the finite dynamic range and noise in the optical intensities causes effective truncation errors. A detailed analysis of finite precision and low-flux photon shot noise is presented in Supplement Section 1.

\begin{figure*}[ht]
\begin{center}
\includegraphics[width=\textwidth]{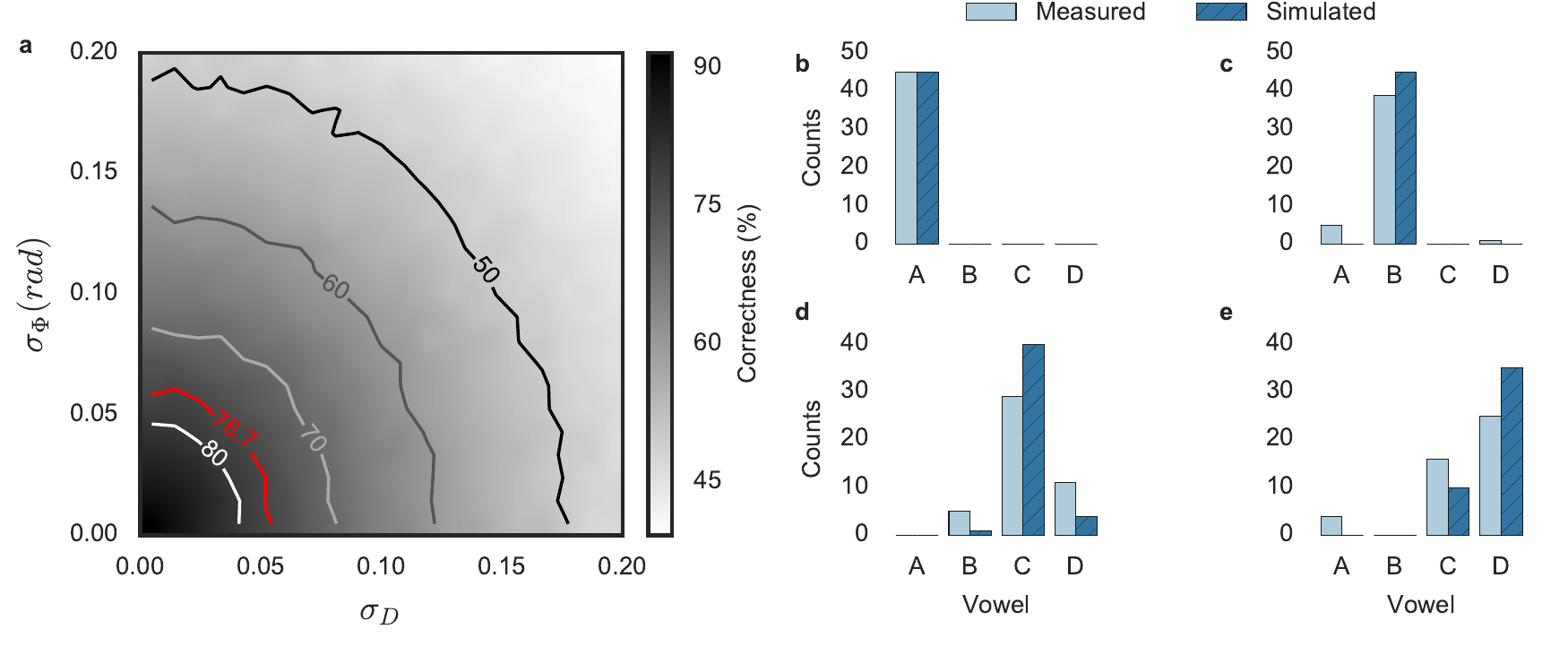}
\caption{\textbf{Vowel recognition.}
(a) Correct rate for vowel recognition problem with various phase encoding error ($\sigma_\Phi$) and photo-detection error ($\sigma_D$), the definition of these two variables can be found in method section. The solid lines are the contours for different level correctness percentage. (b-e) Simulated and experimental vowel recognition results for an error-free training matrix where (b) vowel A was spoken, (c) vowel B was spoken, (d) vowel C was spoken, and (e) vowel D was spoken.
}
\label{fig:vowel_recognition}
\end{center}
\end{figure*}

In this proof-of-concept demonstration, we implement the nonlinear transformation $I_{out}= f(I_{in})$ in the electronic domain, by measuring optical mode output intensities on a photodetector array and injecting signals $I_{out}$ into the next stage of OIU. Here, $f$ models the mathematical function associated with a realistic saturable absorber (such as a dye, semiconductor or graphene saturable absorber or saturable amplifier) that could, in future implementations, be directly integrated into waveguides after each OIU stage of the circuit. For example, graphene layers integrated on nanophotonic waveguides have already been demonstrated as saturable absorbers~\cite{cheng2014plane}. Saturable absorption is modeled as~\cite{selden1967pulse} (Supplement Section 2),
\begin{equation}
\sigma\tau_sI_0=\frac{1}{2}\frac{\ln(T_m/T_0)}{1-T_m},
\label{eqn:nonlinear}
\end{equation}
where $\sigma$ is the absorption cross section, $\tau_s$ is the radiative lifetime of the absorber material, $T_0$ is the initial transmittance (a constant that only depends on the design of saturable absorbers), $I_0$ is the incident intensity, and $T_m$ is the transmittance of the absorber. Given an input intensity $I_0$, one can solve for $T_m(I_0)$ from Eqn.~\ref{eqn:nonlinear}, and the output intensity can be calculated as $I_{out}=I_0\cdot T_m(I_0)$. A plot of the saturable absorber's response function $I_{out}(I_{in})$ is shown in supplement Section 2.

After programming the nanophotonic processor to implement our ONN architecture, which consist of 4 layers of OIUs with 4 neurons on each layer (which requires training a total of $4\cdot6\cdot2=48$ phase shifter settings), we evaluated it on the vowel recognition test set. Our ONN correctly identified 138/180 cases (76.7\%) \cite{endnote1} compared to a simulated correctness of 165/180 (91.7\%).\hl{}

Since our ONN processes information in the analog signal domain, the architecture can be vulnerable to computational errors. Photodetection and phase encoding are the dominant sources of error in the ONN presented here (as discussed above). To understand the role of phase encoding noise and photodection noise in our ONN hardware architecture and to develop a model for its accuracy, we numerically simulate the performance of our trained matrices with varying degrees of phase encoding noise ($\sigma_{\Phi}$) and photodection noise ($\sigma_{D}$) (detailed simulation steps can be found in methods section). The distribution of correctness percentage vs $\sigma_{\Phi}$ and $\sigma_{D}$ is shown in Fig.~\ref{fig:vowel_recognition} (a), which serves as a guide to understanding experimental performance of the ONN. Improvements to the control and readout hardware, including implementing higher precision analog-to-digital converters in the photodetection array and voltage controller, are practical avenues towards approaching the performance of digital computers. Well-known techniques can be applied to engineer the photodiode array to achieve significantly higher dynamic range; for example, using logarithmic or multi-stage gain amplifiers. Addressing these managable engineering problems can further enhance the correctness performance of the ONN to ultimately achieve correctness percentages approaching those of error-corrected digital computers. In addition, ANN parameters trained by conventional back propagation algorithm can become suboptimal when encoding errors are encountered. In such a case, robust simulated annealing algorithms \cite{bertsimas2010robust} can be used to train ANN parameters which is error-tolerant, hence when encoded in the ONN, will have better performance.

\section{Discussion}
Processing \textit{big data} at high speeds and with low power is a central challenge in the field of computer science, and, in fact, a majority of the power and processors in data centers are spent on doing forward propagation (test-time prediction). Furthermore, low forward propagation speeds limit applications of ANNs in many fields including self-driving cars which require high speed and parallel image recognition.

Our optical neural network architecture takes advantage of high detection rate, high-sensitivity photon detectors to enable high-speed, energy-efficient neural networks compared to state-of-the-art electronic computer architectures. Once all parameters have been trained and programmed on the nanophotonic processor, forward propagation computing is performed optically on a passive system. In our implementation, maintaining the phase modulator settings requires some (small) power of $\sim$10 mW per modulator on average. However, in future implementations, the phases could be set with nonvolatile phase-change materials\cite{rios2015integrated}, which would require no power to maintain. With this change, the total power consumption is limited only by the physical size, the spectral bandwidth of dispersive components (THz), and the photo-detection rate (100GHz). In principle, such a system can be at least 2 orders of magnitude faster than electronic neural networks (which are restricted at GHz clock rate). Assuming our ONN has N nodes, implementing $m$ layers of N$\times$N matrix multiplication and operating at a typical 100 GHz photo-detection rate, the number of operations per second of our system would be $$R=2m \cdot N^2\cdot10^{11} \mbox{\mbox{operations/s}}$$

ONN power consumption during computation is dominated by the optical power necessary to trigger an optical nonlinearity and achieve a sufficiently high signal-to-noise ratio (SNR) at the photodetectors (assuming shot-noise limited detection on $n$ photons per pulse, $SNR \simeq \sqrt{1/n}$). We assume a saturable absorber threshold of $p \simeq 1$ MW/cm$^2$ -- valid for many dyes, semiconductors, and graphene \cite{selden1967pulse,Bao2010}. Since the cross section for the waveguide is $A=0.2\upmu$m $\times 0.5\upmu$m, the total power needed to run the system is therefore estimated to be: $P\approx N ~\text{mW}$. Therefore, the energy per operation of ONN will scale as $R/P=2m\cdot N\cdot10^{14}$ operations/J (or $P/R=\frac{5}{mN}$ fJ/operation). Almost the same energy performance and speed can be obtained if optical bistability \cite{soljavcic2002optimal, Nozaki2010, Tanabe:05} is used instead of saturable absorption as the enabling nonlinear phenomenon. Even for very small neural networks, the above power efficiency is already at least 3 orders of magnitude better than that in conventional electronic CPUs and GPUs, where $P/R\approx 1$pJ/operation (not including the power spent on data movement) \cite{horowitz20141}, while conventional image recognition tasks require tens of millions of training parameters and thousands of neurons ($mN\approx10^5$)\cite{krizhevsky2012imagenet}. These considerations suggest that the optical NN approach may be tens of millions times more efficient than conventional computers for standard problem sizes. In fact, the larger the neural network, the bigger the advantage of using optics is: this comes from the fact that evaluating an $N\times N$ matrix in electronics requires $O(N^2)$ energy, while in optics, it requires in principle no energy. Further details on power efficiency calculation can be found in the Supplementary information section 3.

ONNs enable new ways to train ANN parameters. On a conventional computer, parameters are trained with back propagation and gradient descent. However, for certain ANNs where the effective number of parameters substantially exceeds the number of \textit{distinct} parameters (including recurrent neural networks (RNN) and convolutional neural networks(CNN)), training using back propagation is notoriously inefficient. Specifically the recurrent nature of RNNs gives them effectively an extremely deep ANN (depth=sequence length), while in CNNs the same weight parameters are used repeatedly in different parts of an image for extracting features. Here we propose an alternative approach to directly obtain the gradient of each \textit{distinct} parameter without back propagation, using forward propagation on ONN and the finite difference method. It is well known that the gradient for a particular \textit{distinct} weight parameter $\Delta W_{ij}$ in ANN can be obtained with two forward propagation steps that compute $J(W_{ij})$ and $J(W_{ij}+\delta_{ij})$, followed by the evaluation of $\Delta W_{ij}=\frac{J(W_{ij}+\delta_{ij})-J(W_{ij})}{\delta_{ij}}$ (this step only takes two operations). On a conventional computer, this scheme is not favored because forward propagation (evaluating $J(W)$) is computationally expensive. In an ONN, each forward propagation step is computed in \textit{constant} time (limited by the photodetection rate which can exceed 100~GHz~\cite{Vivien:12}), with power consumption that is only proportional to the number of neurons--making the scheme above tractable. Furthermore, with this on-chip training scheme, one can readily parametrize and train unitary matrices--an approach known to be particularly useful for deep neural networks~\cite{arjovsky2015unitary}. As a proof of concept, we carry out the unitary-matrix-on-chip training scheme for our vowel recognition problem (see Supplementary Information Section 4).

Regarding the physical size of the proposed ONN, current technologies are capable of realizing ONNs exceeding the 1000 neuron regime -- photonic circuits with up to 4096 optical components have been demonstrated~\cite{Sun2013}. 3-D photonic integration could enable even larger ONNs by adding another spatial degree of freedom~\cite{rechtsman2013photonic}. Furthermore, by feeding in input signals (e.g. an image) via multiple patches over time (instead of all at once) -- an algorithm that has been increasingly adopted by deep learning community \cite{Jia:Caffe2014} -- the ONN should be able to realize much bigger effective neural networks with relatively small number of physical neurons.

\section{Conclusion}
The proposed architecture could be applied to other artificial neural network algorithms where matrix multiplications and nonlinear activations are heavily used, including convolutional neural networks and recurrent neural networks. Further, the superior forward propagation speed and power efficiency of our ONN can potentially enable training the neural network on the photonics chip directly, using only forward propagation. Finally, it needs to be emphasized that another major portion of power dissipation in current NN architectures is associated with data movement--an outstanding challenge that remains to be addressed. However, recent dramatic improvements in optical interconnects using integrated photonics technology has the potential to significantly reduce data-movement energy cost~\cite{sun2015nature}. Further integration of optical interconnects and optical computing units need to be explored to realize the full advantage of all-optical computing.

\section{Methods}
\textbf{Fidelity Analysis}\\
We evaluated the performance of the $SU(4)$ core with the fidelity metric $f = \sum_i \sqrt{p_i q_i}$
where $p_i, q_i$ are experimental and simulated normalized ($\sum_i x_i = 1$ where $x \in \{p, q\}$) optical intensity distributions across the waveguide modes, respectively.

\textbf{Simulation Method for Noise in ONN}\\
We carry out the following steps to numerically simulate the performance of our trained matrices with varying degrees of phase encoding ($\sigma_{\Phi}$) and detection ($\sigma_{D}$) noise.
\begin{enumerate}
\item For each of the four trained $4\times4$ unitary matrices $U^{k}$, we calculate a set of $\{\theta_i^{k},\phi_i^{k}\}$ that encode the matrix.
\item We add a set of random phase encoding errors, $\{\delta\theta_i^k, \delta\phi_i^k\}$ to the old calculated phases $\{\theta_i^{k},\phi_i^{k}\}$, where we assume each $\delta\theta_i^k$ and $\delta\phi_i^k$ is a random variable sampled from a Gaussian distribution $G(\mu,\sigma)$ with $\mu=0$ and $\sigma=\sigma_{\Phi}$. We obtain a new set of \textit{perturbed} phases $\{{\theta_i^{k}}',{\phi_i^{k}}'\}=\{\theta_i^{k}+\delta\theta_i^k,\phi_i^{k}+\delta\phi_i^k\}$.
\item We encode the four \textit{perturbed} $4\times4$ unitary matrices ${U^{k}}'$ based on the new \textit{perturbed} phases $\{{\theta_i^{k}}',{\phi_i^{k}}'\}$.
\item We carry out the forward propagation algorithm based on the \textit{perturbed} matrices ${U^{k}}'$ with our test data set. During the forward propagation, every time when a matrix multiplication is performed (let's say when we compute $\overrightarrow{v}={U^{k}}'\cdot\overrightarrow{u}$), we add a set of random photo-detection errors $\overrightarrow{\delta v}$ to the resulting $\overrightarrow{v}$, where we assume each entry of $\overrightarrow{\delta v}$ is a random variable sampled from a Gaussian distribution $G(\mu,\sigma)$ with $\mu=0$ and $\sigma=\sigma_{D}\cdot|\overrightarrow{v}|$. We obtain the \textit{perturbed} output vector $\overrightarrow{v}'=\overrightarrow{v}+\overrightarrow{\delta v}$.
\item With the modified forward propagation scheme above, we calculate the correctness percentage for the \textit{perturbed} ONN.
\item Steps 2)-5) are repeated 50 times to obtain the distribution of correctness percentage for each phase encoding noise ($\sigma_{\Phi}$) and photodetection noise ($\sigma_{D}$).
\end{enumerate}

\section{Acknowledgements}
We thank Yann LeCun, Max Tegmark, Isaac Chuang and Vivienne Sze for valuable discussions. This work was supported in part by the Army Research Office through the Institute for Soldier Nanotechnologies under contract W911NF-13-D0001, and in part by the Air Force Office of Scientific Research Multidisciplinary University Research Initiative (FA9550-14-1-0052) and the Air Force Research Laboratory RITA program (FA8750-14-2-0120). M.H. acknowledges support from AFOSR STTR grants, numbers FA9550-12-C-0079 and FA9550-12-C-0038 and Gernot Pomrenke, of AFOSR, for his support of the OpSIS effort, though both a PECASE award (FA9550- 13-1-0027) and funding for OpSIS (FA9550-10-1-0439). N. H. acknowledges support from the National Science Foundation Graduate Research Fellowship grant no. 1122374.

\section{Author Contributions}
Y.S., N.H., S.S., X.S., S.Z., D.E., and M.S., developed the theoretical model for the optical neural network. N.H., M.P., and Y.S. performed the experiment. N.H. developed the cloud-based simulation software for collaborations with the programmable nanophotonic processor. Y.S., S.S., and X.S. prepared the data and developed the code for training MZI parameters. T.B.J. and M.H. fabricated the photonic integrated circuit. All authors contributed in writing the paper.

\bibliography{citations}

\begin{thebibliography}{10}
\bibitem{endnote1}
\bibinfo{note}{On the repeatability of this experiment: we have carried out the
  entire testing run 3 times. The reported result (76.7\% correctness
  percentage) is associated with the best calibrated run (highest measured
  fidelity). The other two less calibrated runs exceeded 70\% correctness
  percentage. For further discussion on enhancing the correctness percentage,
  see S.I. (Section 6).}
\end{thebibliography}
\end{document}